\documentclass[10pt,twoside]{amundi_article}

\usepackage[table]{xcolor}
\usepackage{lmodern}
\usepackage{amssymb}
\usepackage{amsfonts}
\usepackage{amsmath}
\usepackage{amsbsy}
\usepackage{fancybox}
\usepackage{fancyhdr}
\usepackage{graphicx}
\usepackage[hyphens]{url}
\usepackage{arydshln}
\usepackage{multirow}
\usepackage{pdflscape}
\usepackage{tikz}
\usepackage{comment}
\usepackage{nicefrac}
\usepackage{dsfont}
\usepackage{algorithmic}
\usepackage{algorithm}
\usepackage{lipsum}

\newlength{\figurewidth}
\newlength{\figureheight}
\DeclareMathAlphabet\mathbfcal{OMS}{cmsy}{b}{n}
\DeclareFontFamily{OT1}{pzc}{}
\DeclareFontShape{OT1}{pzc}{m}{it}{<-> s * [1.3] pzcmi7t}{}
\DeclareMathAlphabet{\mathpzc}{OT1}{pzc}{m}{it}

\usetikzlibrary{positioning}
\usetikzlibrary{patterns,arrows,decorations.pathreplacing}
\usetikzlibrary{backgrounds}

\usepackage[
        left=3.0cm,
        right=3.0cm,
]{geometry}

\newlength\FHoffset
\fancyheadoffset{\FHoffset}

\usepackage[authoryear]{natbib}

\usepackage{ragged2e}

\usepackage{subfigure}
\usepackage{booktabs}
\definecolor{amundi_blue}{RGB}{0,176,240}
\definecolor{amundi_dark_blue}{RGB}{0,28,75}
\definecolor{darkblue}{rgb}{0.0, 0.0, 0.55}
\usepackage{hyperref}
\hypersetup{
    colorlinks=true,
    citecolor=darkblue,
    linkcolor=darkblue,
    filecolor=darkblue,      
    urlcolor=darkblue
}

\usepackage{pdfpages}
\usepackage{bookmark}

\usepackage{pdfpages}
\usepackage{bookmark}

\usepackage{enumitem}
\usepackage{balance}

\DeclareMathOperator*{\argmin}{arg\,min}

\usepackage{tabularx}

\begin{document}

\setcounter{page}{1}

\title{\textbf{\color{amundi_blue}The Uncertain Shape of Grey Swans: \\
Extreme Value Theory with Uncertain Threshold}%
}

\author{
\hspace{-0.3cm}
{\color{amundi_dark_blue} Hamidreza Arian}\footnote{E-Mail: {\color{amundi_blue} \href{emailto:hamidreza.arian@utoronto.ca}{hamidreza.arian@utoronto.ca}}} \\
\hspace{-0.3cm} Sharif University \\
\hspace{-0.3cm} of Technology
\and
{\color{amundi_dark_blue} Hossein Poorvasei}\footnote{E-Mail: {\color{amundi_blue} \href{emailto:poorvasei@ut.ac.ir}{poorvasei@ut.ac.ir}}} \\
University Tehran
\and
{\color{amundi_dark_blue} Azin Sharifi}\footnote{E-Mail: {\color{amundi_blue} \href{emailto:azin.sharifi@utoronto.ca}{azin.sharifi@utoronto.ca}}} \\
Sharif University \\
of Technology
\and
{\color{amundi_dark_blue} Shiva Zamani}\footnote{E-Mail: {\color{amundi_blue} \href{emailto:zamani@sharif.edu}{zamani@sharif.edu}}} \\
Sharif University \\
of Technology
}

\date{\color{amundi_dark_blue} October 2020}

\maketitle

\begin{abstract}
\noindent
Extreme Value Theory (EVT) is one of the most commonly used approaches in finance for measuring the downside risk of investment portfolios, especially during financial crises. In this paper, we propose a novel approach based on EVT called Uncertain EVT to improve its forecast accuracy and capture the statistical characteristics of risk beyond the EVT threshold. In our framework, the extreme risk threshold, which is commonly assumed a constant, is a dynamic random variable. More precisely, we model and calibrate the EVT threshold by a state-dependent hidden variable, called Break-Even Risk Threshold (BRT), as a function of both risk and ambiguity. We will show that when EVT approach is combined with the unobservable BRT process, the Uncertain EVT's predicted VaR can foresee the risk of large financial losses, outperforms the original EVT approach out-of-sample, and is competitive to well-known VaR models when back-tested for validity and predictability.
\end{abstract}



\section{Introduction}\label{sec:intro}

Extreme Value Theory (EVT) offers insight to model extreme portion of a given general distribution. The early works on EVT include \cite{fisher1928limiting}, \cite{gnedenko1943distribution}, \cite{gumbel1954statistical}, \cite{balkema1974residual} and \cite{pickands1975statistical}. EVT employs two techniques for categorizing extreme events - the Block Maxima and the Peak Over Threshold. The Block Maxima (BM) approach assumes that extreme value data are maxima over certain blocks of time. Peak Over Threshold (POT) method assumes a properly chosen high threshold for extreme events. 
Numerous authors have provided applications of EVT in modelling extreme movements in time series of returns (
\cite{lauridsen2000estimation},  \cite{danielsson2000using}, \cite{danielsson2001using}, \cite{brooks2005comparison}). In addition, after the sub-prime financial crisis, EVT has been used as an ideal framework for modelling fat tail properties of return distributions (\cite{stoyanov2011fat}, \cite{hull2012risk}, \cite{furio2013extreme}). 
EVT can also be used in a multidimensional setting for modeling extreme dependence (\cite{hilal2014portfolio}, \cite{yuen2014upper}). 
Besides from its other advantages, the EVT approach is also capable of using high frequency data for modelling the tail behaviour of asset returns (\cite{bee2016realizing}) and can incorporate price limits in measuring extreme risk (\cite{ji2020combining}). 

Despite the benefits of the EVT framework for risk assessment, the important task of selecting a threshold to separate extreme and non-extreme events poses a great challenge in practice. The choice of the threshold comes before modelling the extreme values, and therefore affects the results of the EVT analysis significantly (\cite{jalal2008predicting}). When the threshold is too high, too few exceedances occur, and when it is too small, the model may not be able to capture the true shape of the tail. One of the earliest studies providing a solution for the threshold selection problem is \cite{dumouchel1983estimating}, suggesting that the threshold can be approximately set as the $95th$ percentile of the empirical distribution. Other approaches can be categorized into graphical approaches based on visual inspection, and analytical approaches of goodness of fit tests. One of the most popular graphical methods commonly used in practice is the Mean Excess Function (MEF) method (\cite{davison1990models}). A major drawback of this approach is that it is subjective and requires human judgment, which makes it hard to implement by a computer and limits its practical applications. As concerned with analytical approaches, some researchers have suggested techniques that provide an optimal trade-off between bias and variance using bootstrap simulations to numerically calculate the optimal threshold (\cite{danielsson2001using}, \cite{drees2000make}, \cite{ferreira2003optimising}, \cite{herrera2013value}, \cite{chukwudum2019optimal}). 

The main contribution of this paper is to propose a novel approach based on the EVT with an application for measuring market risk of financial portfolios. We introduce a state-dependent risk threshold, which we name Break-Even Risk Threshold (BRT), in the EVT framework, based on how risky and ambiguous the stock market is. BRT is estimated such that EVT's risk forecast breaks even with the market's realized future performance. Considering the uncertainty behind an extreme risk threshold, we use variance and ambiguity of return distribution to predict BRT in future periods. The study of \cite{brenner2018asset} introduces ambiguity as a risk-independent variable. Assuming $r_{t+1}$ is the next period's return, they suggest the following risk-ambiguity-return relationship  
\begin{equation}
\mathbb{E}_{t}(r_{t+1}) = r_{f} + \gamma \frac{1}{2} \mathbb{V}{ar}_{t}(r_{t+1})+ \eta\left(1-\mathbb{E}_{t}(\mathrm{P}_{t+1})\right) \mathbb{E}_{t}\left(\left|r_{t+1}-\mathbb{E}_{t}(r_{t+1})\right|\right) \mho_{t}^{2}(r_{t+1}) \nonumber,
\end{equation}
where $\mathrm{P}_{ t+1}$ is the probability of unfavourable returns, $\gamma$ and $\eta$ measure the investor's risk aversion, and sentiment towards ambiguity, respectively. We were inspired by the above relationship to assume that the risk threshold of the EVT, beyond which the tail is modelled, can be a state-dependent variable of risk, as measured by $\mathbb{V}{ar}_{t}$, and ambiguity as measured by $\mho_t^2$. In section \ref{sec:uncertain_evt}, we will talk in more detail about Brenner-Izhakian's measure of ambiguity and the above risk-ambiguity-return relationship. Various authors measure ambiguity in a way that depends on risk and a finite set of moments of the distribution (\cite{epstein2010ambiguity}, \cite{ui2010ambiguity}, \cite{ulrich2013inflation} and \cite{williams2014asymmetric}). However, the ambiguity measure defined by \cite{brenner2018asset} is independent of risk and is calculated using the entire return distribution. In this paper, we use their ambiguity measure along with variance to estimate the risk threshold in the EVT, thereby estimating our so-called Uncertain EVT Value-at-Risk.

In what follows, section \ref{sec:glance_evt} provides an overview of the topic of EVT for modelling the tail of return distributions. Emphasizing the importance of risk threshold, in section \ref{sec:BRT}, we introduce the novel concept of BRT. In section \ref{sec:uncertain_evt}, we discuss sources of uncertainty in the stock market and provide an application of risk and ambiguity in estimating BRT. Finally, in section \ref{sec:emp}, we provide numerical results on six major global indices to show the efficiency of predicting VaR using BRT and compare our results to some other well-known approaches. Section \ref{sec:con} concludes the paper.
\section{A Glance at Extreme Value Theory}\label{sec:glance_evt}

In this section, we start with a general introduction to Extreme Value Theory (EVT) and its applications in measuring market risk. Various measures of risk have been introduced by academics and practitioners in the past several decades. One such measure is Value-at-Risk (VaR), formally defined as
\begin{equation}
\mathbb{P}\left(X_T<-\text{VaR}_p\right) = 1-p, 
\end{equation}
where $X_T$ is a portfolio's return for the time horizon $T$ and $p$ is the confidence level for measuring risk. 

Extreme Value Theory (EVT) is generally used with Peak Over Threshold (POT) method (\cite{mcneil1997peaks},\cite{mcneil2000estimation}, \cite{genccay2004extreme}). The POT method considers observations exceeding a high threshold, sometimes called grey swans, and models these occurrences separately from the rest of the return distribution. Here grey swans, in contrast to Nassim Taleb's highly improbable black swans (\cite{taleb2007black}), are financial loss scenarios beyond a risk threshold $u$, which have a low probability of occurrence but are still possible. EVT is concerned with such events in the tail of profit and loss distribution. The conditional distribution function of observations beyond the threshold, $F_u(x)$, is 
\begin{equation}
F_{u}(x) = \mathbb{P}(X-u \leq x | X>u) = \frac{F(u+x)-F(u)}{1-F(u)} ,
\end{equation}
where $F(u)$ is the original cumulative distribution function. For an appropriate threshold $u$, $F_u(x)$ can be approximated by the Generalized Pareto Distribution (GPD), which follows the form 
\begin{equation}\label{eq:gpd}
{G_{\xi ,\sigma ,u}}(x) = \left\{
\begin{array}{lr}
1 - (1 + \xi \frac{{x - u}}{\sigma})^{-\frac{1}{\xi}}  \quad & \text{if} \xi \ne 0 , \\
1 - e^{-\frac{x-u}{\sigma}} \quad  & \text{if} \xi =0 ,
\end{array}\right.
\end{equation}
with the shape and scale parameters $\xi$  and $ \sigma $ , respectively, and $u$ is the risk threshold (see \cite{balkema1974residual} and \cite{pickands1975statistical}). $\xi$ determines the possible shape of a grey swan with values of $\xi > 0$ corresponding to the heavy-tailed distributions. For the case of non-zero $\xi$, the density function of GPD, as defined by equation \eqref{eq:gpd}, is given by
\begin{equation}\label{eq:gpd_density}
g_{{\sigma },\xi}({x}) = \frac{1}{{{\sigma }}}\,{\left(1 + {\xi }\frac{x-u}{{{\sigma }}}\right)^{ - \frac{1}{{{\xi }}} - 1}}. 
\end{equation}

There are three parameters which need to be estimated to use EVT for calculating VaR; $u$, $\sigma$ and $\xi$. Arguably, choosing the appropriate risk threshold is the most challenging part of the model calibration. For tackling the threshold estimation problem, classical approaches, as discussed in section \ref{sec:intro}, rely on tuning parameters that the practitioner selects manually. Various authors in the past two decades have proposed alternative solutions for the problem of finding an appropriate threshold $u$ to provide a balance between low and high threshold estimates. \cite{scarrott2012review} provides a comprehensive review of some recent and classical techniques with an emphasis on the uncertainty involved in the threshold estimation algorithms. Considering threshold uncertainty, \cite{behrens2004bayesian}, proposes a mixture model for the middle and the tail of the distribution. As another example of a recent work for choosing a suitable threshold, \cite{attalides2015threshold} develops a Bayesian inference method taking into account the uncertainty in threshold selection. Their Bayesian cross-validation approach uses an average estimation from different threshold estimates. \cite{bader2016automated} proposes an automatic threshold selection algorithm by choosing the lowest goodness-of-fit of the tail distribution to the exceedances. Highlighting the importance of the automated threshold selection, \cite{schneider2019threshold} introduces two data-driven threshold selection procedures by evaluating the variance of log-spacings from the exponential distribution. 

Even though estimating risk threshold $u$ is challenging in its nature, estimating the shape and scale parameters is straightforward. Once $u$ is set, calibrating parameters $\xi$ and $\sigma$ is easily done using Maximum Likelihood Estimation (MLE). After estimating the risk threshold and the GPD parameters, the VaR of the underlying return distribution is calculated by
\begin{equation}\label{5}
\text{VaR} = u + \frac{{\hat \sigma }}{{\hat \xi }}\left\{ {{{\left[ {\frac{n}{{{n_u}}}(1 - p)} \right]}^{ - \hat\xi }} - 1} \right\},
\end{equation}
with $n \text{ and } n_u$ being the size of the sample and the number of observations above $u$, respectively. 

\section{Break-Even Risk Threshold (BRT)}\label{sec:BRT}

\begin{figure}{ht!}
\centering
\includegraphics[width=15cm]{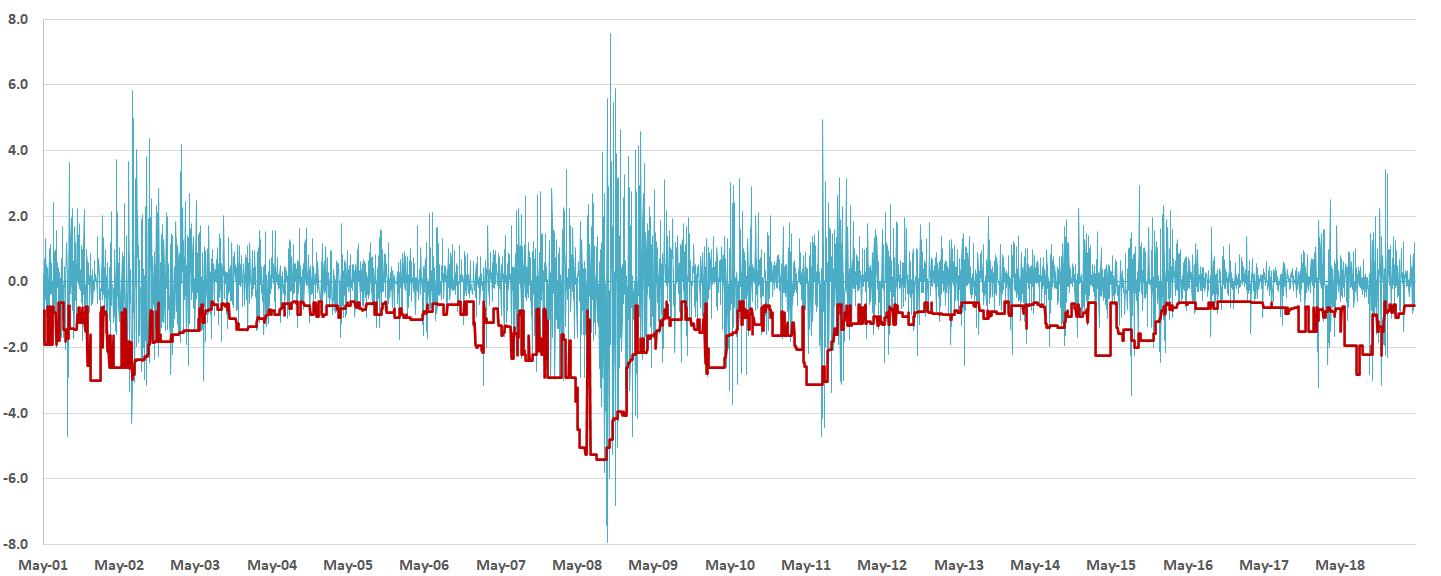}
\caption{The time series of realized Break-Even Risk Threshold (BRT) for S\&P 500 returns (in red) calculated from equation \eqref{eq:BRT_definition1}. The horizontal axis represents time and the vertical axis shows return and BRT values (in percentage).}
\label{fig:actual_brt_sp500}%
\end{figure}

Assume $ \text{VaR}^\text{EVT}_p (T_1, t; u)$ is the Value-at-Risk at confidence level $p$ calculated by EVT method with threshold $u$ using input data from time $T_1$ to current time $t$, where $T_1 < t$. Let us also imagine that we are able to see the future state of the market return, and let $\text{VaR}^\text{H}_p (t+1, T_2)$ be the historical Value-at-Risk at confidence level $p$ using input data from time $t+1$ to $T_2$, where $t +1\leq T_2$. We define the BRT at current time $t$ as the value of $u_t$ in a domain $\mathcal{D}$ such that the VaR calculated by EVT replicates the historical VaR based on \textit{future} data. In mathematical terms, we define $\text{BRT}_t$, as

\begin{equation}\label{eq:BRT_definition1}
    \text{BRT}_t^\text{Realized} = \argmin_{\hat{u}\in\mathcal{D}}\left|\text{VaR}^\text{EVT}_p (T_1, t; \hat{u}) - \text{VaR}^\text{H}_p (t+1, T_2)\right|,
\end{equation}
where $\mathcal{D} \subset \mathbb{R}^-$ refers to a domain where the minimum is taken. For an efficient and fast estimation of BRT, in equation \eqref{eq:BRT_definition1}, we limit the search space $\mathcal{D}$ to the negative realized returns from time $T_1$ to $t$. In the above definition, as ironic as it sounds, the \textit{historical} VaR is calculated based on \textit{future} data to estimate risk. In mathematical terms, we search for a threshold $u$ such that
\begin{equation}
\text{VaR}^\text{EVT}_p (T_1, t; u) \approx \text{VaR}^\text{H}_p (t+1, T_2).
\end{equation}
Since we certainly can not see the future state of the market, we try to find relevant information which can be used to recover BRT without using future data. Figure \ref{fig:actual_brt_sp500} shows the realized BRT calculated for the S\&P 500 index using equation \eqref{eq:BRT_definition1}. As it is clear from the figure, during the financial crisis of 2007-08, the BRT has dramatically changed to a very extreme regime.

When it comes to managing risk of large-scale financial portfolios, regulators and risk managers have diverse sets of concerns and preferences (\cite{christoffersen2001testing}). A VaR method reflecting concerns of regulators on risk measurement forecasts must guarantee few violations for a given level of confidence (\cite{christoffersen2004backtesting}). Risk managers, apart from considering regulatory expectations, must consider their firm's profitability and business growth by choosing less conservative measures of risk. \cite{Rossignolo2012} highly support EVT compared to its competitors for risk measurement by financial institutions. They argue that risk measurement using EVT would have protected banks from massive losses and the consequent economic capital required during the 2008 crisis. Considering their findings, a powerful risk measure should be able to provide flexibility in meeting risk manager's preferences. Fortunately, by simply adjusting the hyper-parameters of BRT, one can design a VaR engine under the EVT framework to meet risk manager's needs. For instance, in equation \eqref{eq:BRT_definition1}, if we set the time-frame for calculating historical forward-looking VaR to one business day, $T_2=t+1$, the BRT satisfies 

\begin{equation}\label{eq:BRT_definition2}
    \text{BRT}_t^\text{Realized}  = \argmin_{\hat{u}\in\mathcal{D}}\left|\text{VaR}^\text{EVT}_p (T_1, t; \hat{u}) - r_{t+1}\right|,
\end{equation}
where $r_{t+1}$ is the return for the next business day. This way, we are able to track return time series and better utilize capital under management while it is more likely to violate predetermined VaR confidence levels. Whereas using equation \eqref{eq:BRT_definition1}, implies that the violation ratio of EVT matches that of the realized VaR. In sections \ref{sec:model_validation} and \ref{sec:model_predictability}, we will show the numerical implications of using various time windows, $[t+1, T_2]$ in the definition of BRT, and its impact on the final VaR measure.
\begin{figure}{ht!}
\centering
\subfigure[high ambiguity]{{\includegraphics[width=6.8cm]{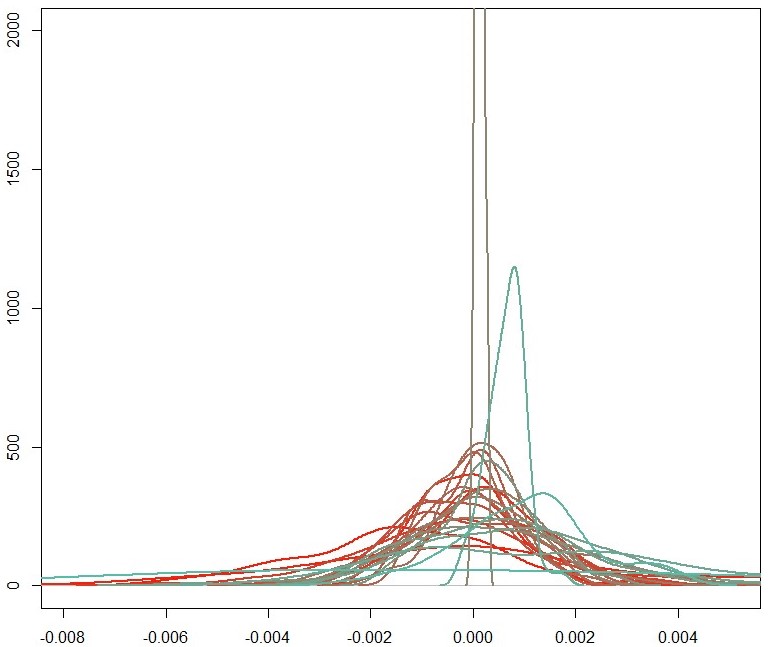}}}
\quad
\subfigure[low ambiguity]{{\includegraphics[width=6.9cm]{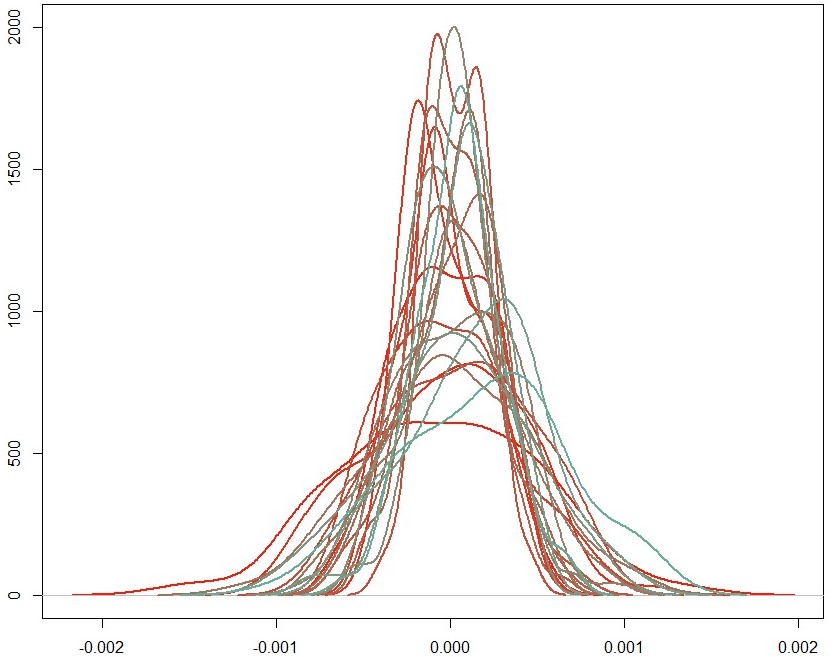}}}
\caption{In this figure, panel a) shows daily distributions of returns for a month with relatively high ambiguity where the horizontal and vertical axes show return and density, respectively. In panel b) a sample of daily returns for a month with low ambiguity is displayed.}\label{fig:motivation}%
\end{figure}

\section{Extreme Value Theory with Uncertain Threshold}\label{sec:uncertain_evt}

In this section, we explain the concept of ambiguity and its relationship with risk and return, paving the way to link uncertainty and EVT. Uncertainty in the stock market can be decomposed into risk and ambiguity components (\cite{brenner2018asset}). The key to an appropriate empirical measure of ambiguity is its separation from financial risk estimates. Considering this separation, we attempt to model risk and ambiguity independently and use them together to estimate the BRT.

To define ambiguity, we first introduce some mathematical notations. Assume a probability space $(\Omega, \mathbb{P}, \mathcal{F})$ with $\sigma-$algebra $\mathcal{F}$, and probability measure $\mathbb{P}$ on which the intraday return process $r$ is defined. Moreover, suppose $\mathcal{P}$ is a set of possible probabilities of stock returns on which we can define a probability measure $\mu$. Now we assume the intraday return $r$ has an unknown probability $\varphi(r)$ with probability $\mu$ on the set of probability measures $\mathcal{P}$, then the expected marginal probability of return and its variance on $\mathcal{P}$ are 
\begin{equation}
\mathbb{E}[\varphi(r)] \equiv \int_{\mathcal{P}} \varphi(r) d \mu, \qquad \operatorname{\mathbb{V}ar}[\varphi(r)] \equiv \int_{\mathcal{P}}(\varphi(r)-\mathbb{E}[\varphi(r)])^{2} d \mu,    
\end{equation}
respectively.

Following \cite{brenner2018asset}, ambiguity can be measured by
\begin{equation}\label{eq:ambiguity_definition}
\mho^{2}[r]=\int \operatorname{\mathbb{V}ar}[\varphi(r)] {\mathbb{E}}[\varphi(r)] dr.
\end{equation}
In the above equation, ${\mathbb{E}}[\varphi(r)] $ is a density function of expected distribution probabilities under measure $\mu$ and therefore, $\mho^2[r]$ reflects the expected variance of probability distributions, $\operatorname{\mathbb{V}ar} [\varphi(r)] $, of intraday equity returns. 

Regarding the difference between the concepts of risk and ambiguity, we emphasize that risk deals with known probability distributions over future outcomes, whereas ambiguity refers to situations where the probability distribution is unknown. From this subtle difference, aversion to ambiguity indicates that individuals prefer known probabilities, and they are willing to pay to avoid ambiguous market events. Most studies on ambiguity do not provide managerial insight for using ambiguity for the purpose of financial decision making. Moreover, only a limited number of studies use market data to measure ambiguity (see \cite{ulrich2013inflation, williams2014asymmetric}). In this paper, we aim to highlight the role that risk and ambiguity play in identifying extreme events.

Selecting a suitable threshold for fitting GPD on excess returns is a critical step in the EVT approach. Traditional techniques suggest constant risk threshold, but in this paper, we introduce an unobservable, dynamic and state-dependent risk threshold that evolves over time. The two factors we are using to predict a risk threshold are variance and ambiguity. These two parameters can be used to explain possible financial scenarios beyond an extreme risk threshold $u$, which have a low probability of occurrence. When the variance of the return distribution is higher, there are more sample returns away from zero. Therefore in times of volatile markets, we need a threshold away from zero to better construct the tail of the return distribution.  

Less intuitively, high ambiguity levels suggest that there is more fear built into the stock market than traditional volatility measures indicate. When ambiguity is high, there is a high level of uncertainty in market participants' behaviour, and investors need to be compensated for future market chaos. Figure \ref{fig:motivation}, panel a, shows S\&P 500 intraday returns distribution in a month with high ambiguity. As we can see, there is a high degree of dispersion between distributions. We expect that in times of high ambiguity (future unrest or times of high dispersion between intraday return distributions), the optimal level of risk threshold $u$ is closer to zero. Figure \ref{fig:motivation}, panel b, demonstrates intraday returns distribution in a month with low ambiguity. It is evident that dispersion between these distributions has a less degree of uncertainty, and therefore we expect less room for deviations of risk threshold as well. Intuitively, we expect to have a negative relationship between BRT and ambiguity. If ambiguity of original distribution is high, then the ambiguity of GPD, representing the tail of the original distribution, is expected to be high, as well. Therefore, we are more uncertain about the location of the GPD for future time periods. By approaching u to zero, we can address this uncertainty and provide more flexibility for the GPD. 

It is a well known stylized fact in asset pricing theory that risk as compared with ambiguity has a stronger and more visible impact on the equity premium. When \textit{ambiguity} premium is added alongside \textit{risk} premium, it forms equity \textit{uncertainty} premium. The case of ambiguity premium is more involved, however, and depends on the probability of favourable returns. When this probability is high (low), the premium on ambiguity is positive (negative).

To compute the ambiguity measure numerically, \cite{brenner2018asset} assumes $\varphi(r) = \phi(r; \mu, \sigma)$ is a normal probability density function with mean $\mu$ and standard deviation $\sigma$, and calculates ambiguity on a monthly basis by each day's return probability variance  
\begin{equation}
\mho^{2}[r]=\int \mathbb{E}[\phi(r ; \mu, \sigma)] \operatorname{\mathbb{V}ar}[\phi(r ; \mu, \sigma)] d r ,
\end{equation}
then they use the following approximation, based on normality assumption, to assess the degree of monthly ambiguity

\begin{equation}
\mho^{2}[r] = \sum_{i=1}^N {\frac{1}{w_i(1-w_i)} {\operatorname{\mathbb{V}ar}\left[\Pi_i\right]} \mathbb{E}\left[\Pi_i\right]},  
\end{equation}
where $N$ is the number of normal histogram bins on which the empirical daily return distribution is fit, for the $i$th bin, the bin size is $w_i$ and the term $\frac{1}{w_i(1-w_i)}$ is a scale factor. The probabilities $\Pi_i$ are computed from normal cumulative probability functions as
\begin{equation}
\Pi_i = \Phi(r_{i}; \boldsymbol{\mu}, \boldsymbol{\sigma}) - \Phi(r_{i-1}; \boldsymbol{\mu}, \boldsymbol{\sigma}),
\end{equation}
with $\Phi(r_{0}; \boldsymbol{\mu}, \boldsymbol{\sigma}) = 0$ and $\Phi(r_{N+1}; \boldsymbol{\mu}, \boldsymbol{\sigma}) = 1$. Assuming 21 days in the valuation month, the vectors $\boldsymbol{\mu} \text{ and } \boldsymbol{\sigma}$ contain the means and standard deviations of normal distributions fitted to each day's data. The expectation operator, $\mathbb{E}$, and variance operator, $\mathbb{Var}$, compute the mean and variance of probabilities $\Pi_i$ across valuation month given the mean and standard deviation vectors $\boldsymbol{\mu}$ and $\boldsymbol{\sigma}$. Contrary to \cite{brenner2018asset}, our bin size is not constant and depends on how far the bin is away from zero. For returns in $[-2\%, 2\%]$, our intervals are $0.1\%$ wide. For other ranges $[-3\%, -2\%] \cup [2\%, 3\%]$, $[-4\%, -3\%] \cup [3\%, 4\%]$, $[-5\%, -4\%] \cup [4\%, 5\%]$, $[-6\%, -5\%] \cup [5\%, 6\%]$, the bin sizes increase progressively as $0.2\%$, $0.25\%$, $0.5\%$, and $1\%$, respectively. Moreover, instead of normal histogram probabilities, we use the percentage of actual occurrences per bin. 
An interesting result by \cite{bi2020value} shows a negative relationship between Value-at-Risk and expected return. On the expected return forecasting, \cite{brenner2018asset} proposes the risk-ambiguity-return relationship as
\begin{equation}\label{eq:risk_return_ambiguity}
\mathbb{E}_{t}(r_{t+1}) = r_{f} + \gamma \frac{1}{2} \mathbb{V}{ar}_{t}(r_{t+1}) + \eta\left(1-\mathbb{E}_{t}(\mathrm{P}_{t+1})\right) \mathbb{E}_{t}\left(\left|r_{t+1}-\mathbb{E}_{t}(r_{t+1})\right|\right) \mho_{t}^{2}(r_{t+1}),
\end{equation}
where $\mathbb{E}_{t}(r_{t+1})$ is next business day's expected return, $r_f$ is the risk free rate and the second and third terms on the right hand side represent risk and ambiguity premiums, respectively, $\gamma$ measures the investor's risk aversion, and $\eta$ measures investor's sentiment towards ambiguity, which depends on the expected probability of favourable returns $(1-\mathbb{E}_{t}(\mathrm{P}_{ t+1}))$. Motivated by the above relationship, we introduce a random and uncertain risk threshold for EVT, which reflects the investor's expectations as illustrated by both risk and ambiguity. Therefore, we could model the threshold using risk and ambiguity with a multiple linear regression
\begin{equation}\label{eq:brt_regression}
    \text{BRT}_t = \beta_{0} + \beta_{1}\sigma_{t-21}^{2} + \beta_{2}\mho_{t-21}^{2},
\end{equation}
where $\text{BRT}_t$ is the risk threshold at time $t$, $\sigma_{t-21}^{2}$ is the 21-day historical variance and $\mho_{t-21}^{2}$ is the previous month level of ambiguity. 21 days is selected because ambiguity is calculated monthly using intraday return data. In this case, VaR can be calculated by 
\begin{equation}\label{eq:var_uncertain_evt}
    \text{VaR}_t = \text{BRT}_t + \frac{{\hat \sigma }(\text{BRT}_t)}{{\hat \xi }(\text{BRT}_t)}\left\{ {{{\left[ {\frac{n}{{{n_u}}}(1 - {p})} \right]}^{ - {\hat \xi}(\text{BRT}_t) }} - 1} \right\},
\end{equation}
Note that parameters $\hat\xi$ and $\hat\sigma$ are functions of threshold $\text{BRT}_t$ and therefore the shape and scale of the tail of distribution are affected by the level of risk and ambiguity in the underlying portfolio. 

\section{Empirical Analysis}\label{sec:emp}

In this section, we provide a detailed description of the market data we have used for putting our improvement of the EVT-VaR approach into test. We provide numerical results on the estimation of our dynamic threshold model with ambiguity and variance. We compute VaR for our selected indices and then back-test our results using some well-known approaches.

\subsection{Data Description}

In this paper, six major global indices including S\&P 500 (USA), FTSE 100 (UK), Dow Jones (USA), Nikkei (Japan), BVSP (Brazil) and Merval (Argentina) from April 2005 \footnote{Starting dates for some of these indices are earlier than April 2005 (see Figure \ref{fig:Uncertain_VaR}).} until October 2019 are analyzed. To forecast BRT, we use 5-minute return data of indices to calculate ambiguity on a monthly basis, and daily closing prices to estimate variance. We use data from Trade and Quote (TAQ) and Finam databases. The reasons behind selecting these indices are that, first, they form a diverse set of developed as well as emerging markets; second, they represent some of the important stock markets in the world, and many funds replicate them as their investment portfolios. 

To better understand the data, Table \ref{tab:tabstat} represents skewness, kurtosis, maximum, minimum, and Jarque-Bera test results for daily stock returns. The data we use follows fat-tail skewed distributions, and Jarque-Bera test indicates that the returns are not normal (\cite{giot2004modelling}).

\subsection{Forecasting BRT}

To forecast BRT, we use a rolling window of 600 days, $[T, T+599]$, as a training period to estimate BRT dynamics and the next 25 days,$[T+600, T+624]$, as the test period to forecast BRT. Inside the 600-day training window, we choose two rolling windows. First, a rolling window, $[T_1, t]$, of 100 days to compute $ \text{VaR}^\text{EVT}_p (T_1, t; u)$, and second, a rolling window, $[t+1, T_2]$, of 50 days to compute $\text{VaR}^\text{H}_p (t+1, T_2)$. We choose $\mathcal{D}$ in equation \eqref{eq:BRT_definition1} to be all the negative returns in the interval $[T_1, t]$. Our algorithm searches for the optimal $u$ in the search space $\mathcal{D}$ satisfying equation \eqref{eq:BRT_definition1}.

\begin{figure}
\centering
\subfigure[S\&P 500]{{\includegraphics[width=7cm]{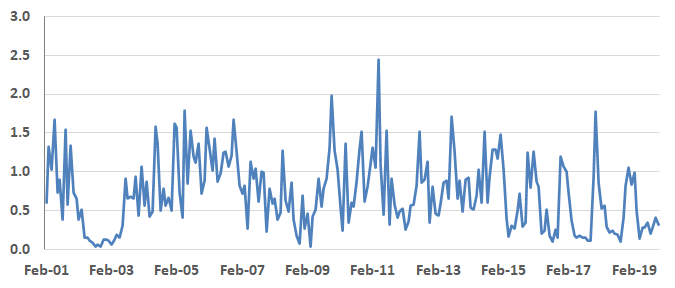}}}
\quad
\subfigure[FTSE 100]{{\includegraphics[ width=7cm]{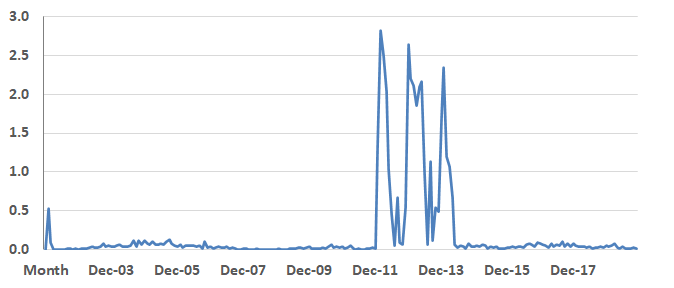}}}
\\
\subfigure[Dow Jones]{{\includegraphics[ width=7cm]{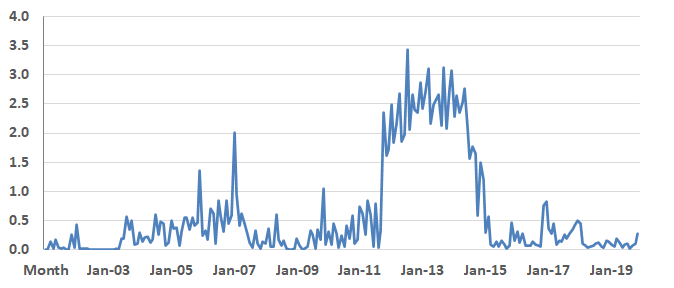}}}
\quad
\subfigure[Nikkei]{{\includegraphics[ width=7cm]{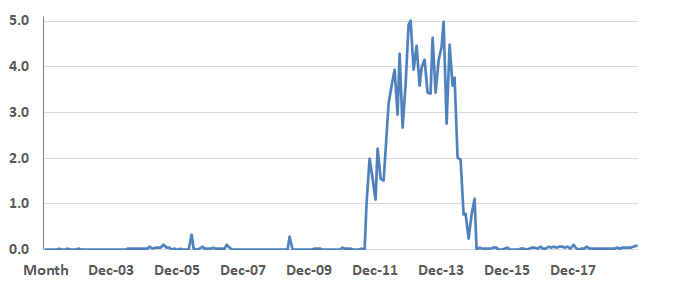}}}
\\
\subfigure[BVSP]{{\includegraphics[ width=7cm]{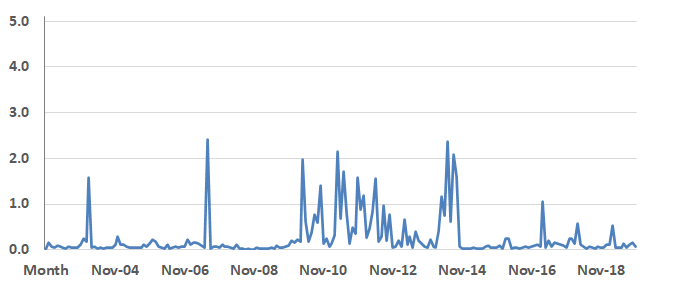}}}
\quad
\subfigure[Merval]{{\includegraphics[ width=7cm]{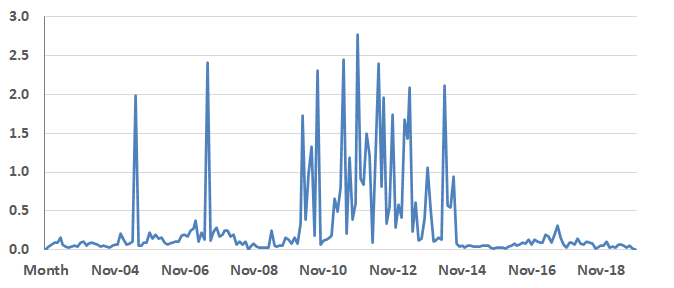}}}
\caption{Time series of ambiguity for six major global indices: S\&P 500, FTSE 100, Dow Jones, Nikkei, BVSP and Merval. The horizontal axis represents time and the vertical axis represents ambiguity.} \label{fig:ambiguity}%
\end{figure}

In the next step, we fit the linear regression \eqref{eq:brt_regression} on the calculated BRTs, as the response variable, against two independent variables, one month historical variance and ambiguity, on rolling window $[T+100, T+549]$. We calculate ambiguity for mentioned indices based on 5-minute return data.\footnote{Our R code for computing ambiguity is available upon request.} Figure \ref{fig:ambiguity} demonstrates ambiguity time series of these indices. Note that the regression equation \eqref{eq:brt_regression} cannot be estimated on the entire 600-day rolling window, $[T, T+599]$, as the first 100 days are allocated to the first training interval and the last 50 days are allocated to the last training interval. In our regressions, the independent variables are significant predictors of BRTs.

\begin{table}
  \centering
  \caption{Descriptive statistics of the daily return data}
    \begin{tabular}{lccccc}
    \toprule
          & skewness & kurtosis & maximum & minimum & Jarque Bera \\
    \midrule
    S\&P 500 & -0.35 & 10.58 & 7.57 & -7.93 & R \\
    FTSE 100 & 0.07 & 10.84 & 7.95 & -7.81 & R \\
    Dow Jones & -0.21 & 11.17 & 7.77 & -7.05 & R \\
    Nikkei & -0.40 & 16.99 & 12.36 & -10.00 & R \\
    BVSP  & 0.11 & 10.03 & 14.45 & -11.11 & R \\
    Merval & -1.47 & 28.54 & 10.14 & -31.65 & R \\
    \bottomrule
    \end{tabular}
  \label{tab:tabstat}%
  \vspace{1ex}\\
     {\raggedright Note: This table represents statistical properties of the data used in this paper and results of Jarque-Bera (JB) normality test. Min and max returns are in percentage and for the JB test, R means that the normality hypothesis is rejected. Kurtosis results indicate that all the return time series are fat-tailed. \par}
\end{table}%

\subsection{VaR Estimation}

\begin{figure*}
   \centering
    \includegraphics[scale=0.45]{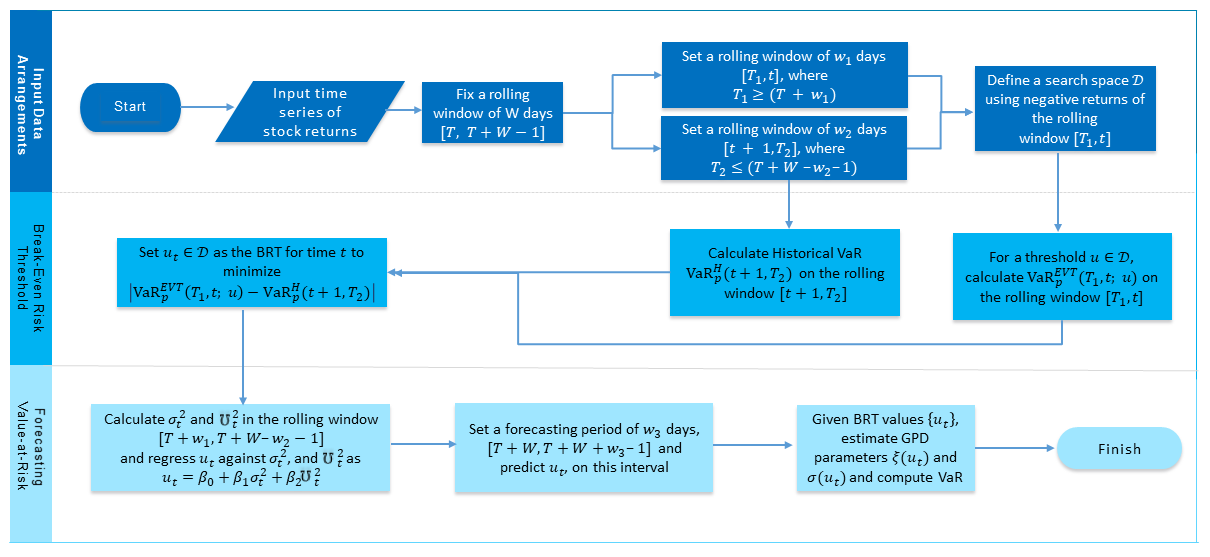}
    \caption{This figure presents the flowchart of our Uncertain EVT model. In input data arrangements, the valuation time is $T$, the rolling window size $W=600$. $\text{VaR}^\text{EVT}_p (T_1, t; u)$ is computed on the interval $[T_1, t]$, with the rolling window size of $w_1 = 100$. $\text{VaR}^\text{H}_p (t+1, T_2)$ is computed on $[t+1, T_2]$ with the rolling window size of $w_2 = 50$. Finally, assuming forecasting window size of $w_3=25$, the BRT process is predicted on the interval $[T+600, T+624]$. }
   \label{fig:uncertain_evt_flowchart}
\end{figure*}

\begin{figure}[ht!]
\centering
\subfigure[S\&P 500]{{\includegraphics[width=7cm]{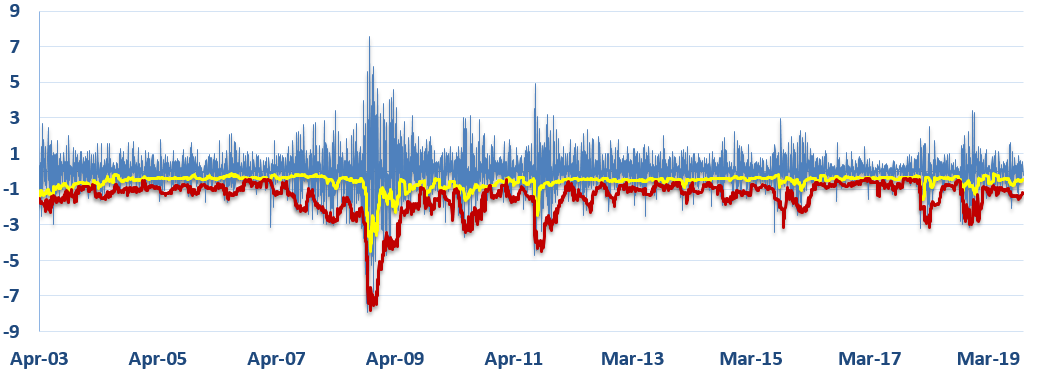}}}
\quad
\subfigure[FTSE 100]{{\includegraphics[width=7cm]{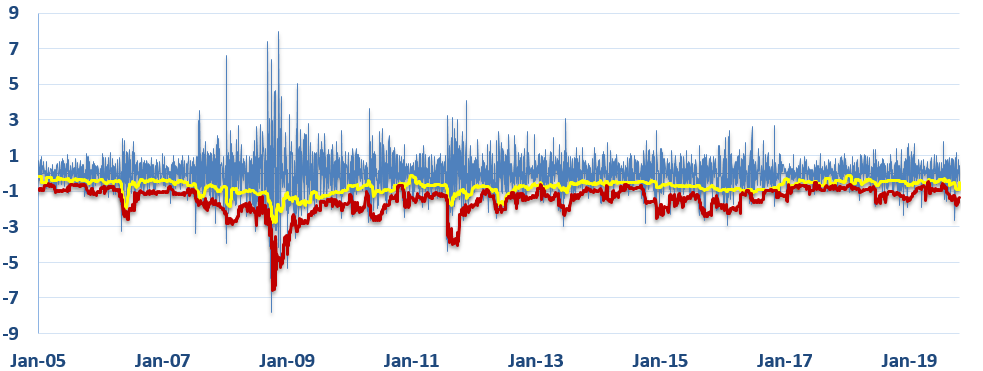}}}
\\
\subfigure[Dow Jones]{{\includegraphics[width=7cm]{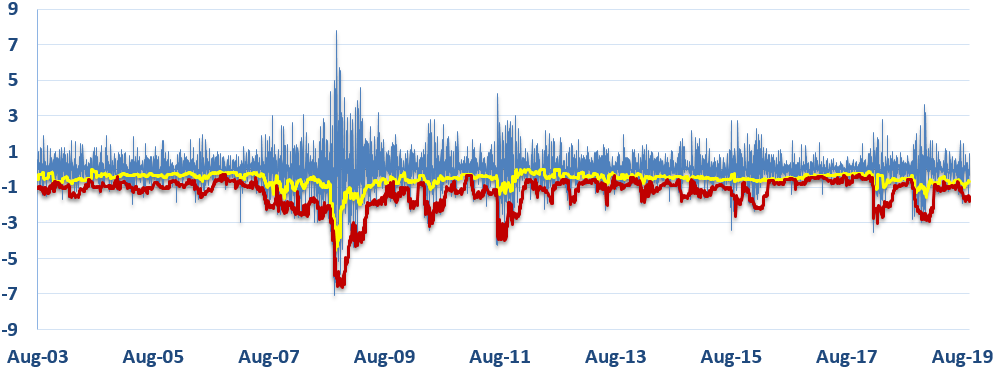}}}
\quad
\subfigure[Nikkei]{{\includegraphics[width=7cm]{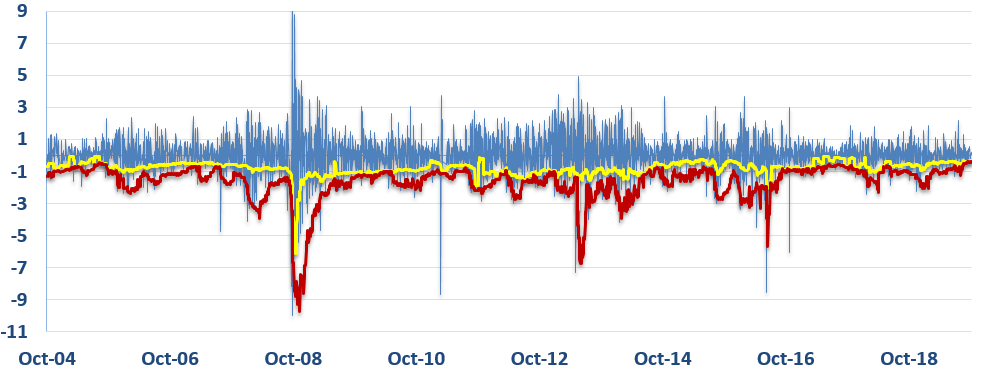}}}
\\
\subfigure[BVSP]{{\includegraphics[width=7cm]{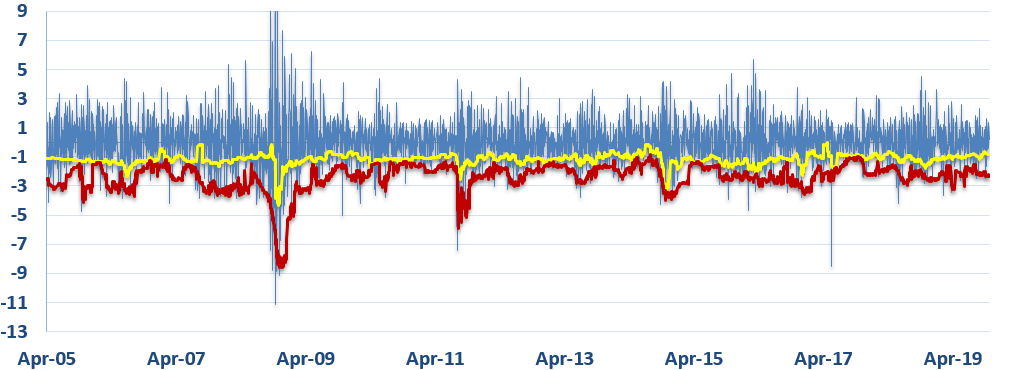}}}
\quad
\subfigure[Merval]{{\includegraphics[width=7cm]{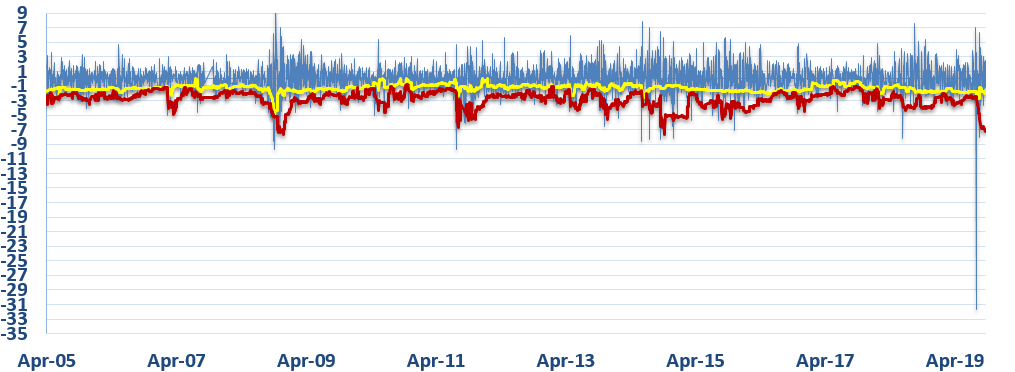}}}
\caption{VaR results (in red) using Uncertain EVT approach with $95\%$ confidence level, using equation \eqref{eq:var_uncertain_evt}, as well as predicted BRTs (in yellow), using equation \eqref{eq:brt_regression}, are displayed for S\&P 500, FTSE 100, Dow Jones, Nikkei, BVSP and Merval. The horizontal and vertical axes represent time and return, respectively.} 
\label{fig:Uncertain_VaR}%
\end{figure}

Using the fitted regression model from the previous step, we predict the BRT for the time interval $[T+600, T+624]$. Once the threshold is estimated, historical returns below the threshold are used to find GPD parameters, $\xi$ and $\sigma$, in equation \eqref{eq:gpd} via Maximum Likelihood Estimation (MLE). Finally, using equation \eqref{eq:var_uncertain_evt}, we estimate daily VaR with confidence level $95\%$ for the next 25 days. Figure \ref{fig:uncertain_evt_flowchart} summarizes all the aforementioned steps in calculating VaR. As the results of our approach, Figure \ref{fig:Uncertain_VaR} presents the time series of predicted BRT and the Uncertain EVT VaR for six major indices in our analysis.

\begin{figure*}
   \centering
\subfigure[Uncertain EVT]{{\includegraphics[width=7cm]{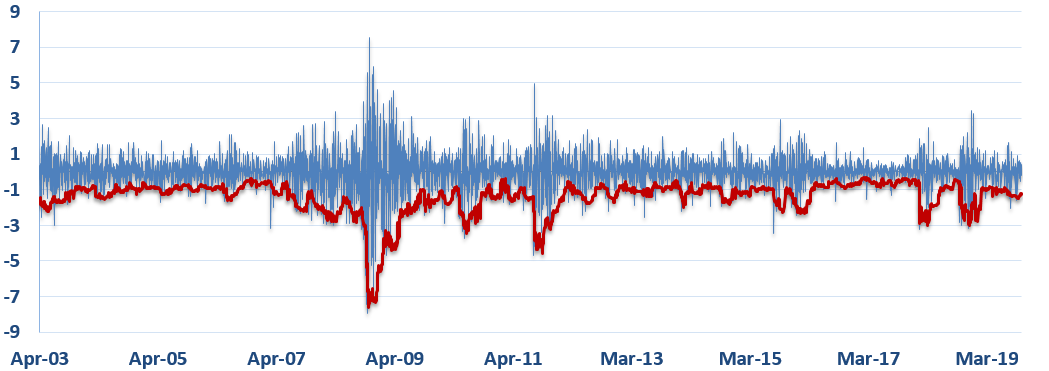}}}
\quad
\subfigure[EVT]{{\includegraphics[width=7cm]{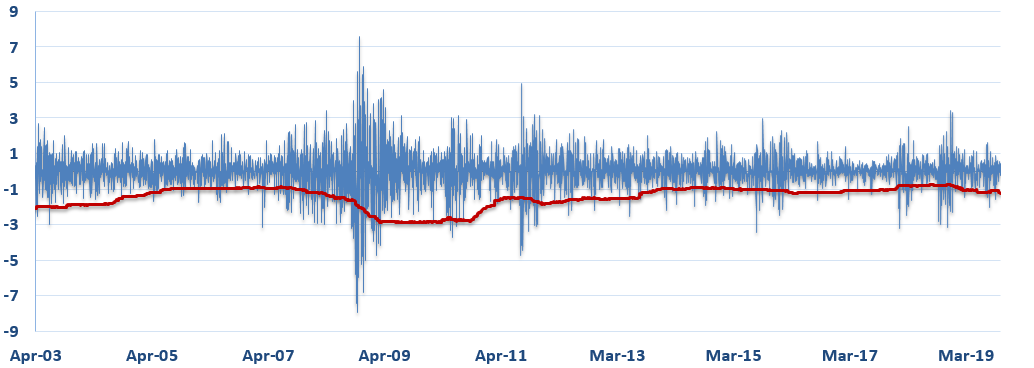}}}
\\
\subfigure[EGARCH]{{\includegraphics[width=7cm]{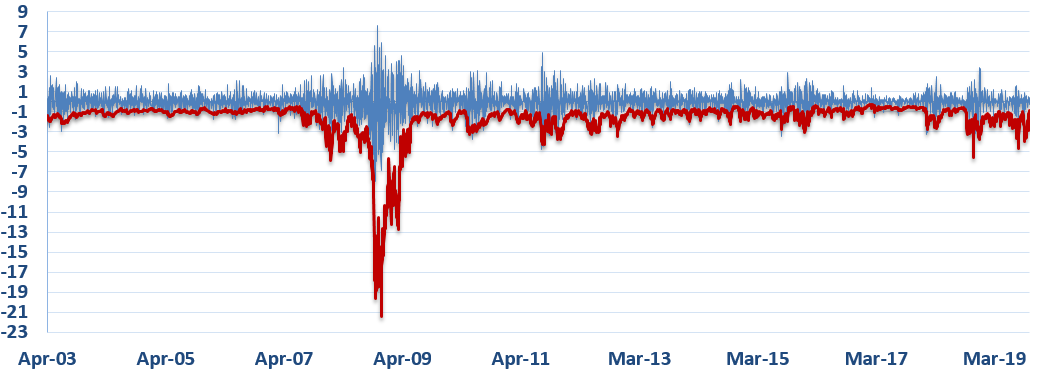}}}
\quad
\subfigure[GARCH]{{\includegraphics[width=7cm]{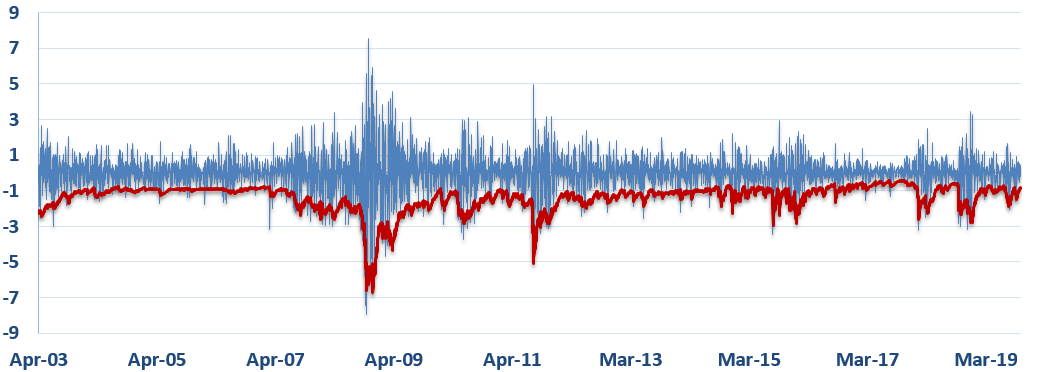}}}
\\
\subfigure[CaviaR asymmetric]{{\includegraphics[width=7cm]{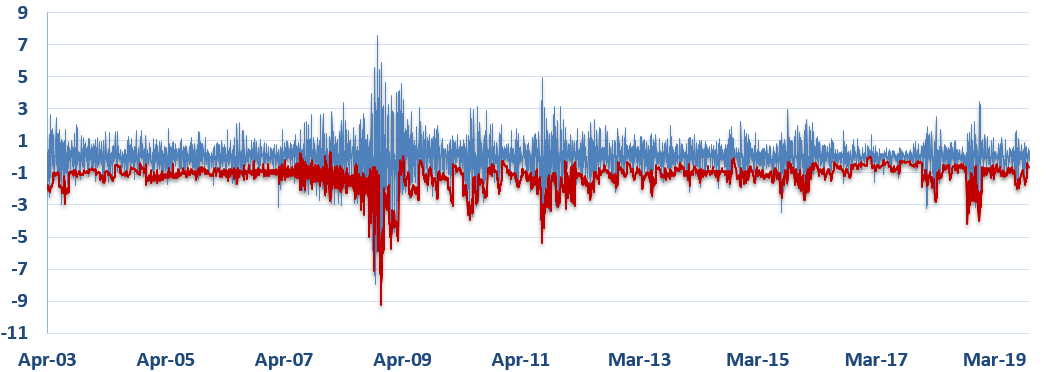}}}
\quad
\subfigure[Monte Carlo Simulation]{{\includegraphics[width=7cm]{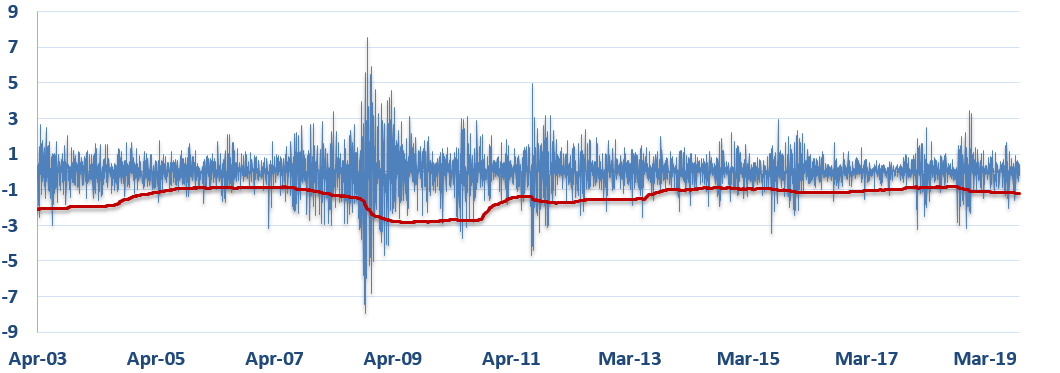}}}
\\
\subfigure[Historical Simulation]{{\includegraphics[width=7cm]{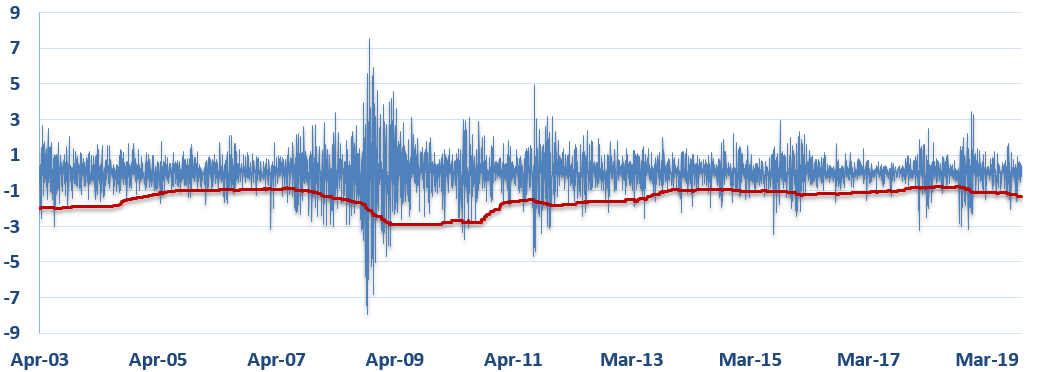}}}
\quad
\subfigure[Variance-Covariance]{{\includegraphics[width=7cm]{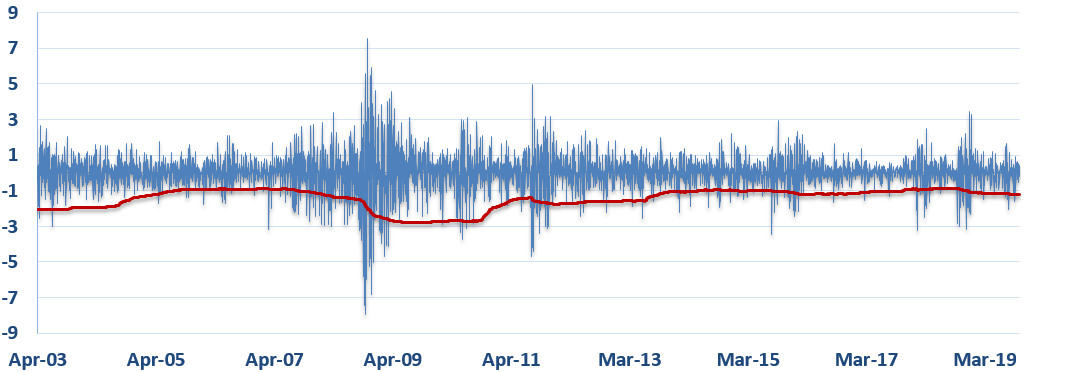}}}
    \caption{
    VaR results (in red) with $95\%$ confidence level using various methods including Uncertain EVT, EVT, EGARCH, GARCH, CaviaR asymmetric, Monte Carlo Simulation, Historical Simulation, and Variance-Covariance methods for S\&P 500 are displayed. The horizontal and vertical axes represent time and return, respectively.}
\label{fig:var_benchmark_comparison}
\end{figure*}

\subsection{Model Validation}\label{sec:model_validation}

Two of the most common methods used for VaR back-testing are unconditional and conditional coverage methods (\cite{kupiec1995techniques}, \cite{christoffersen1998evaluating}). Unconditional coverage method concentrates only on the number of violations, whereas conditional coverage method also considers the dependence of violations in successive periods. For more details about these two tests see Appendix \ref{sec:appendix_backtesting}. In Table \ref{tab:ModelValidation} the performance of our VaR model, Uncertain EVT, is compared with seven other approaches, including EVT, EGARCH, GARCH, CaviaR asymmetric, Monte Carlo Simulation, Historical Simulation, and Variance-Covariance methods. For a brief overview of benchmark methods in our comparative analysis, see Appendix \ref{sec:appendix_benchmark}. A competitive approach is GARCH, where none of its VaR results is rejected except for the FTSE 100 index under the unconditional coverage test. Among all the indices, our model's results are not rejected except for the Merval index under conditional coverage test. Overall, the results of back-testing shows a strong performance of Uncertain EVT {in which our method has improved the results of the EVT approach.}

\begin{table}
  \centering
  \caption{Back-test results for VaR with $95\%$ confidence level using likelihood ratio unconditional and conditional coverage tests of Kupiec and Christoffersen, respectively.}
\resizebox{\textwidth}{!}{
    \begin{tabular}{lccccccccccc}
\cmidrule{1-11}    \multirow{2}[4]{*}{Method} & \multicolumn{1}{c}{\multirow{2}[4]{*}{Backtest Method}} & \multicolumn{3}{c}{S\&P 500} & \multicolumn{3}{c}{FTSE 100} & \multicolumn{3}{c}{Dow Jones} &  \\
\cmidrule{3-11}          &       & t stat & p-value &       & t stat & p-value &       & t stat & p-value &       &  \\
\cmidrule{1-11}    \multirow{2}[1]{*}{Uncertain EVT} & LR UC & 0.883 & 0.347 & NR    & 1.382 & 0.239 & NR    & 2.834 & 0.092 & NR    &  \\
          & LR CC & 1.035 & 0.595 & NR    & 5.777 & 0.055 & NR    & 3.571 & 0.167 & NR    &  \\
    \multirow{2}[0]{*}{EVT} & LR UC & 0.188  & 0.664 & NR     & 3.854  & 0.049 & R     & 2.484  & 0.114 & NR     &  \\
          & LR CC & 22.6  & 1.19E-05 & R     & 10.838  & 0.004 & R     & 18.2  & 0.0001 & R     &  \\
    \multirow{2}[0]{*}{EGARCH} & LR UC & 1.110 & 0.292 & NR    & 12.313 & 0.00045 & R     & 13.936 & 0.00019 & R     &  \\
          & LR CC & 2.065 & 0.356 & NR    & 12.639 & 0.00180 & R     & 19.216 & 0.00007 & R     &  \\
    \multirow{2}[0]{*}{GARCH} & LR UC & 0.032 & 0.857 & NR    & 5.377 & 0.020 & R     & 1.540 & 0.215 & NR    &  \\
          & LR CC & 2.414 & 0.299 & NR    & 5.460 & 0.065 & NR    & 4.438 & 0.109 & NR    &  \\
\multirow{2}[0]{*}{CaviaR asymmetric} & LR UC & 31.2  & 2.38E-08 & R     & 5.1   & 2.36E-02 & R     & 15.8  & 7.09E-05 & R \\
          & LR CC & 31.3  & 1.60E-07 & R     & 9.0   & 1.11E-02 & R     & 16.1  & 3.22E-04 & R \\
    \multirow{2}[0]{*}{Monte Carlo Simulation} & LR UC & 3.160 & 0.075 & NR    & 4.140 & 0.042 & R     & 4.343 & 0.037 & R     &  \\
          & LR CC & 15.152 & 0.001 & R     & 15.381 & 0.000 & R     & 25.062 & 0.000 & R     &  \\
    \multirow{2}[0]{*}{Historical Simulation} & LR UC & 0.424 & 0.515 & NR    & 4.434 & 0.035 & R     & 2.485 & 0.115 & NR    &  \\
          & LR CC & 15.212 & 0.000 & R     & 12.376 & 0.002 & R     & 20.046 & 4.44E-05 & R     &  \\
    \multirow{2}[1]{*}{Variance-Covariance} & LR UC & 0.616 & 0.433 & NR    & 3.854 & 4.96E-02 & R     & 1.421 & 0.233 & NR    &  \\
          & LR CC & 16.684 & 2.38E-04     & R     & 18.894 & 7.89E-05 & R     & 20.826 & 0.00003 & R     &  \\
\cmidrule{1-11}    \multirow{2}[4]{*}{Method} & \multicolumn{1}{c}{\multirow{2}[4]{*}{Backtest Method}} & \multicolumn{3}{c}{Nikkei} & \multicolumn{3}{c}{BVSP} & \multicolumn{3}{c}{Merval} &  \\
\cmidrule{3-11}          &       & t stat & p-value &       & t stat & p-value &       & t stat & p-value &       &  \\
\cmidrule{1-11}    \multirow{2}[1]{*}{Uncertain EVT} & LR UC & 2.824 & 0.092 & NR    & 0.653 & 0.418 & NR    & 0.107 & 0.743 & NR    &  \\
          & LR CC & 2.974 & 0.225 & NR    & 1.570 & 0.455 & NR    & 14.720 & 0.0006 & R     &  \\
    \multirow{2}[0]{*}{EVT} & LR UC & 0.226  & 0.634 & NR     & 0.350  & 0.553 & NR     & 1.877  & 0.170 & NR     &  \\
          & LR CC & 4.966  & 0.083 & NR     & 6.348  & 0.041 & R     & 24.1  & 5.91E-06 & R     &  \\
    \multirow{2}[0]{*}{EGARCH} & LR UC & 6.783 & 0.00920 & R     & 0.270 & 0.603 & NR    & 0.009 & 0.926 & NR    &  \\
          & LR CC & 9.461 & 0.00882 & R     & 2.789 & 0.248 & NR    & 0.299 & 0.861 & NR    &  \\
    \multirow{2}[0]{*}{GARCH} & LR UC & 1.891 & 0.169 & NR    & 0.004 & 0.950 & NR    & 0.028 & 0.867 & NR    &  \\
          & LR CC & 4.556 & 0.102 & NR    & 0.013 & 0.993 & NR    & 2.345 & 0.310 & NR    &  \\
    \multirow{2}[0]{*}{CaviaR asymmetric} & LR UC & 3.562 & 0.059 & NR    & 1.298 & 0.255 & NR    & 7.398 & 0.007 & R \\
          & LR CC & 3.580 & 0.167 & NR    & 6.340 & 0.042 & R     & 7.992 & 0.018 & R \\
    \multirow{2}[0]{*}{Monte Carlo Simulation} & LR UC & 2.331 & 0.127 & NR    & 0.025 & 0.875 & NR    & 2.990 & 0.084 & NR    &  \\
          & LR CC & 4.445 & 0.108 & NR    & 8.721 & 0.013 & R     & 17.303 & 0.000 & R     &  \\
    \multirow{2}[0]{*}{Historical Simulation} & LR UC & 0.107 & 0.743 & NR    & 0.653 & 0.419 & NR    & 2.515 & 0.113 & NR    &  \\
          & LR CC & 3.372 & 0.185 & NR    & 6.112 & 0.047 & R     & 23.553 & 0.000 & R     &  \\
    \multirow{2}[1]{*}{Variance-Covariance} & LR UC & 3.355 & 0.067 & NR    & 0.185 & 0.667 & NR    & 1.878 & 0.171 & NR    &  \\
          & LR CC & 4.954 & 0.084 & NR    & 7.491 & 0.024 & R     & 17.761 & 0.00014 & R     &  \\
\cmidrule{1-11}
\end{tabular}%
}
  \label{tab:ModelValidation}%
\vspace{1ex}\\
     {\justifying \noindent Note: This table presents back-testing results of various VaR methods, including Uncertain EVT, EVT, GARCH, EGARCH, CaviaR asymmetric, Monte Carlo Simulation, Historical Simulation, and Variance-Covariance using unconditional and conditional coverage tests. Unconditional coverage test only cares about the number of violations not to exceed a predetermined confidence level, but conditional coverage test also considers successive violations. In this table, LR UC and LR CC refer to likelihood ratio unconditional and conditional coverage tests, respectively. As per the results of these coverage tests, NR stands for not rejected, and R stands for rejected results. Two of the most successful methods based on the results on global indices S\&P 500, FTSE 100, Dow Jones, Nikkei, BVSP and Merval are the Uncertain EVT and GARCH methods. 
     \par}
\end{table}

\subsection{Model Predictability}\label{sec:model_predictability}

Apart from back-testing results, in this study, we employ another test to compare the predictive ability of our benchmark approaches to Uncertain EVT. \cite{diebold2002comparing} provides a popular approach to compare the prediction power of two given models. A detailed discussion on Diebold-Mariano predictive ability test is given in Appendix \ref{sec:appendix_backtesting}. Taking into account risk managers' concerns, we use equation\eqref{eq:BRT_definition2} to calculate the corresponding BRT and compare its predictive ability to other benchmark models. Test results for global indices are shown in Tables \ref{tabDM_SP_FTSE_DJ} and \ref{tabDMNikkei_BVSP_Merval}.

Looking at these tables as matrices, the $ij$th entry of Diebold-Mariano test statistics provides the predictive ability of model $i$ versus model $j$. When this number is less (more) than the critical value $-1.64$ ($+1.64$), we conclude the model $i$ is significantly superior (inferior) to model $j$.

As we can see, Uncertain EVT shows a strong performance with respect to other benchmarks. Among all the indices, we observe that Uncertain EVT has a moderate performance in FTSE 100 and the second strongest performance in the rest. In FTSE 100 index, GARCH, EGARCH and CaviaR asymmetric have performed better than Uncertain EVT. {One thing to be noticed is that Uncertain EVT method outperforms the EVT approach in all the indices through which one can conclude that our method has a stronger predictive power than the method it is originated from.} Figure 
\ref{fig:var_benchmark_comparison} provides the time series of S\&P 500 historical returns as well as VaR using eight different approaches, including our Uncertain EVT method.

\begin{table}
  \centering
  \caption{Diebold-Mariano predictive ability test results for S\&P 500, FTSE 100 and Dow Jones indices at $95\%$ confidence level.}
\resizebox{\textwidth}{!}{
    \begin{tabular}{l|cccccccc}
          \toprule
    \multicolumn{1}{c}{\multirow{2}[4]{*}{DM test stat}} & \multicolumn{8}{c}{S\&P 500} \\
\cmidrule{2-9}    \multicolumn{1}{c}{} & Uncertain EVT & MCS   & HS    & VC    & EGARCH & GARCH & EVT   & CaviaR a \\
    \midrule
    Uncertain EVT & -     & -15.24 & -19.01 & -17.88 & -9.32 & -2.16 & -17.62 & 8.76 \\
    MCS   &       & -     & -26.18 & -35.19 & 11.79 & 9.83  & -16.90 & 16.78 \\
    HS    &       &       & -     & 1.89  & 13.87 & 16.78 & 13.22 & 21.75 \\
    VC    &       &       &       & -     & 14.85 & 18.33 & 3.29  & 22.43 \\
    EGARCH &       &       &       &       & -     & 1.62  & -13.04 & 16.49 \\
    GARCH &       &       &       &       &       & -     & -15.92 & 15.18 \\
    EVT   &       &       &       &       &       &       & -     & 20.74 \\
    CaviaR a &       &       &       &       &       &       &       & - \\
    \midrule
    \multicolumn{1}{c}{\multirow{2}[4]{*}{DM test stat}} & \multicolumn{8}{c}{FTSE 100} \\
\cmidrule{2-9}    \multicolumn{1}{c}{} & Uncertain EVT & MCS   & HS    & VC    & EGARCH & GARCH & EVT   & CaviaR a \\
    \midrule
    Uncertain EVT & -     & -6.20 & -7.24 & -6.50 & 14.02 & 11.22 & -7.80 & 5.55 \\
    MCS   &       & -     & -9.82 & 11.15 & 17.92 & 20.66 & -8.12 & 11.05 \\
    HS    &       &       & -     & 2.62  & 20.14 & 22.22 & -23.33 & 13.07 \\
    VC    &       &       &       & -     & 18.58 & 20.59 & -4.64 & 11.44 \\
    EGARCH &       &       &       &       & -     & -5.58 & -20.99 & -6.69 \\
    GARCH &       &       &       &       &       & -     & -22.84 & -4.22 \\
    EVT   &       &       &       &       &       &       & -     & 14.47 \\
    CaviaR a &       &       &       &       &       &       &       & - \\
    \midrule
    \multicolumn{1}{c}{\multirow{2}[4]{*}{DM test stat}} & \multicolumn{8}{c}{Dow Jones} \\
\cmidrule{2-9}    \multicolumn{1}{c}{} & Uncertain EVT & MCS   & HS    & VC    & EGARCH & GARCH & EVT   & CaviaR a \\
    \midrule
    Uncertain EVT & -     & -10.37 & -12.93 & -13.25 & -14.17 & -6.91 & -9.82 & 11.95 \\
    MCS   &       & -     & -16.27 & -45.85 & 3.23  & 4.38  & -15.09 & 17.48 \\
    HS    &       &       & -     & 8.76  & 5.70  & 10.66 & -4.49 & 22.93 \\
    VC    &       &       &       & -     & 5.56  & 8.29  & -2.22 & 21.31 \\
    EGARCH &       &       &       &       & -     & 7.69  & -5.59 & 28.14 \\
    GARCH &       &       &       &       &       & -     & -10.51 & 22.47 \\
    EVT   &       &       &       &       &       &       & -     & 20.25 \\
    CaviaR a &       &       &       &       &       &       &       & - \\
    \bottomrule
    \end{tabular}%
    }
  \label{tabDM_SP_FTSE_DJ}%
\vspace{1ex}\\
     {\justifying \noindent Note: The first row compares the predictive ability performance of Uncertain EVT versus EVT, EGARCH, GARCH, Monte Carlo Simulation (MCS), Historical Simulation (HS), Variance-Covariance (VC) and CaviaR asymmetric (CaviaR a) approaches. It can be concluded for S\&P 500 and Dow Jones indices, Uncertain EVT shows stronger predictive performance compared to all the methods except for CaviaR asymmetric. For FTSE 100, Uncertain EVT performs better except for EGARCH, GARCH, and CaviaR asymmetric approaches.
     \par}
\end{table}%

\begin{table}
  \centering
  \caption{Diebold-Mariano predictive ability test results for Nikkei, BVSP and Merval indices at $95\%$ confidence level.}
    \resizebox{\textwidth}{!}{
    \begin{tabular}{l|cccccccc}
\toprule
    \multicolumn{1}{c}{\multirow{2}[4]{*}{DM test stat}} & \multicolumn{8}{c}{Nikkei} \\
\cmidrule{2-9}    \multicolumn{1}{c}{} & Uncertain EVT & MCS   & HS    & VC    & EGARCH & GARCH & EVT   & CaviaR a \\
    \midrule
    Uncertain EVT & -     & -13.93 & -9.62 & -14.58 & -12.74 & -12.77 & -10.23 & 4.07 \\
    MCS   &       & -     & 12.61 & -0.56 & 5.40  & 8.75  & 19.73 & 14.58 \\
    HS    &       &       & -     & -23.53 & -1.69 & 2.75  & -18.76 & 8.58 \\
    VC    &       &       &       & -     & 5.56  & 9.55  & 22.73 & 18.44 \\
    EGARCH &       &       &       &       & -     & 5.98  & -0.21 & 17.83 \\
    GARCH &       &       &       &       &       & -     & -4.88 & 15.22 \\
    EVT   &       &       &       &       &       &       & -     & 9.20 \\
    CaviaR a &       &       &       &       &       &       &       & - \\
    \midrule
    \multicolumn{1}{c}{\multirow{2}[4]{*}{DM test stat}} & \multicolumn{8}{c}{BVSP} \\
\cmidrule{2-9}    \multicolumn{1}{c}{} & Uncertain EVT & MCS   & HS    & VC    & EGARCH & GARCH & EVT   & CaviaR a \\
    \midrule
    Uncertain EVT & -     & -8.33 & -4.32 & -9.67 & -0.67 & 0.38  & -5.80 & 5.60 \\
    MCS   &       & -     & 32.63 & -22.74 & 12.14 & 12.59 & 30.58 & 12.97 \\
    HS    &       &       & -     & -37.82 & 8.26  & 9.22  & -8.97 & 7.21 \\
    VC    &       &       &       & -     & 14.09 & 14.57 & 33.34 & 15.76 \\
    EGARCH &       &       &       &       & -     & -3.62 & -9.54 & 6.34 \\
    GARCH &       &       &       &       &       & -     & -10.44 & 7.43 \\
    EVT   &       &       &       &       &       &       & -     & 8.52 \\
    CaviaR a &       &       &       &       &       &       &       & - \\
    \midrule
    \multicolumn{1}{c}{\multirow{2}[4]{*}{DM test stat}} & \multicolumn{8}{c}{Merval} \\
\cmidrule{2-9}    \multicolumn{1}{c}{} & Uncertain EVT & MCS   & HS    & VC    & EGARCH & GARCH & EVT   & CaviaR a \\
    \midrule
    Uncertain EVT & -     & -5.48 & -8.19 & -7.05 & -0.83 & 0.31  & -8.25 & 4.90 \\
    MCS   &       & -     & -15.45 & -16.59 & 5.19  & 7.11  & -20.46 & 14.96 \\
    HS    &       &       & -     & 4.02  & 8.84  & 10.07 & -18.08 & 17.20 \\
    VC    &       &       &       & -     & 8.61  & 10.11 & -9.46 & 17.37 \\
    EGARCH &       &       &       &       & -     & 1.87  & -9.10 & 10.50 \\
    GARCH &       &       &       &       &       & -     & -10.92 & 8.48 \\
    EVT   &       &       &       &       &       &       & -     & 17.85 \\
    CaviaR a &       &       &       &       &       &       &       & - \\
    \bottomrule
    \end{tabular}%
    }
  \label{tabDMNikkei_BVSP_Merval}%
\vspace{1ex}\\
     {\justifying \noindent Note: The first row compares the predictive ability performance of Uncertain EVT versus other approaches. It can be concluded for Nikkei, BVSP and Merval indices that except for CaviaR asymmetric, Uncertain EVT has stronger predictive performance compared to EVT, EGARCH, GARCH, Monte Carlo (MCS), Historical Simulation (HS) and Variance-Covariance (VC) approaches, which for BVSP and Merval the performance of Uncertain EVT versus GARCH and EGARCH does not show a significant difference.
     \par}
\end{table}%

\subsection{BRT and Tail Estimation During Crisis Periods}

One of the innovations of this paper is the introduction of BRT as an unobservable latent variable. As it is clear from Figure \ref{fig:actual_brt_sp500}, during the crisis period of December 2007 to June 2009, the actual BRT process dramatically drops. Calculation of the actual BRT process for other indices also shows similar behaviour of BRT during market turbulence. Figure \ref{fig:Uncertain_VaR}, shows that the forecasted BRT process, also sharply decreases during financial crises.

This is important to note that, in the EVT framework, there are two distributions we are dealing with while measuring risk of financial portfolios, the original distribution, and the GPD. Estimation of the change in behaviour of the latter is at the center of this research. During financial crises, we observe more extreme deviations from the mean, and therefore, it does make sense to choose a lower risk threshold to discern the tail from the rest of the original distribution. The advantage of this lower threshold is that it enables us to fit the GPD more realistically, and leads us to a more accurate VaR.

\section{Conclusions and Future Research}\label{sec:con}

We presented a novel approach based on Extreme Value Theory for estimating VaR where the threshold beyond which a GPD is modelled as the tail of the distribution is not a constant but a state-dependent variable depending on both variance and ambiguity. The combined effect of variance and ambiguity, which is often referred to as uncertainty, is strongly affecting the optimal level of threshold. Numerous cases show that our approach, the Uncertain EVT, improves the predictability of the EVT approach and is competitive to some of the most advanced and efficient VaR methods developed so far. 

Several advantages of our model are as follows. First, instead of using historical methods for calculating the Extreme Risk Threshold, we proposed an economically meaningful technique to predict the extreme level beyond which the tail should be modelled. Second, The dynamic nature of our approach helps improve the accuracy and robustness of VaR estimation when moving into and out of crisis periods. This is important as financial models are criticized for 

Third, the approach we offer is flexible to be used by risk managers who are interested in obtaining a risk measure meeting certain back-testing criterion, such as violation ratios or loss functions. 

For future research, we point out that there might be factors other than variance and ambiguity, explaining the dynamic behaviour of BRT. Another approach might model BRT as an autoregressive process of its own lagged values and previous returns. Apart form factors affecting BRT, there might be other modelling frameworks to predict the next state of an optimal EVT threshold. The BRT time series, as Augmented Dickey-Fuller test indicates, shows a strong mean-reversion property. Therefore, it also might be a good idea to model BRT directly as a stochastic mean-reverting process. 

\clearpage

\setcitestyle{numbers} 
\bibliographystyle{plainnat}
\bibliography{refs}

\clearpage

\appendix
\section*{Appendix}
\section{An Overview of Benchmark Models}\label{sec:appendix_benchmark}

In this section, we provide a brief overview of all the benchmark methods used in this paper. We divide all VaR methods into three categories: non-parametric, parametric and semi-parametric. Non-parametric approaches assume no parametric distribution for the returns and try to extract the distribution from historical data by different techniques. In parametric approaches, simple parametric distributions, like normal and student's-t are assumed for the returns. Semi-parametric approaches combine different techniques of parametric and non-parametric approaches.

\subsection{Non-parametric Methods}

\subsubsection{Historical Simulation}
This method uses a rolling window in historical data and estimates the experimental distribution of the losses, then the one period ahead VaR is calculated as a specific quantile of this distribution.

\subsubsection{Monte Carlo Simulation }
Monte Carlo simulation method simulates future returns based on an explicit formula and then implement historical simulation method on that data to calculate one period ahead VaR. In this paper, we have used Geometric Brownian Motion (GBM) to simulate the price of an asset,  $S_{t}$, as
\begin{equation}
    d S_{t}=\mu S_{t} d t+\sigma S_{t} d W_{t},
\end{equation}
where constants $\mu$ and $\sigma$ are called drift and diffusion, respectively. $W_{t}$ is the Wiener process with $W_{t} \sim \mathcal{N}(0,t)$. 

\subsection{Parametric Methods}

\subsubsection{Variance-Covariance}
In this method, a rolling window is used, and the standard deviation of returns from this window is calculated. Assuming normal returns with mean zero, one can measure VaR at time $t$ using
\begin{equation}
\label{eq:var_covar}
    VaR_{t} = \mathcal{N}^{-1}(\Theta)\sigma_{t},
\end{equation}
where $\mathcal{N}^{-1}$ is the inverse of cumulative standard normal distribution and $\theta$ is a specific confidence level.

\subsubsection{GARCH}
Generalized Autoregressive Conditional Heteroskedasticity (GARCH) model tries to forecast future variances of return series using lagged variances and returns. In GARCH(p,q) model we have

\begin{equation}
    \varepsilon_{t}=\sqrt{\sigma_{t}} \eta_{t}, \quad \eta_{t}^{\mathrm{IID}} \sim \mathcal{N}(0,1),
\end{equation}
\begin{equation}
    \sigma_{t}^2=\alpha_{0}+\sum_{i=1}^{q} \alpha_{i} \varepsilon_{t-i}^{2}+\sum_{j=1}^{p} \beta_{j} \sigma_{t-j}^2,
\end{equation}
where $\alpha$ and $\beta$ are constants,  $\varepsilon_{t}$ is the error term and $\sigma_{t}$ is the variance of $\varepsilon_{t}$ conditional on the information available up to time $t$. Then we use equation \ref{eq:var_covar} to calculate VaR. One can use student's-t distribution instead of normal distribution for $\eta_{t}$.

\subsubsection{EGARCH}
EGARCH model is an extension to GARCH model which better depicts volatility asymmetry in financial data. In this model, we have
\begin{equation}
    \log \sigma_{t}^{2}=\omega+\sum_{k=1}^{q} \beta_{k} g\left(Z_{t-k}\right)+\sum_{k=1}^{p} \alpha_{k} \log \sigma_{t-k}^{2},
\end{equation}
where $g\left(Z_{t}\right)=\theta Z_{t}+\lambda\left(\left|Z_{t}\right|-E\left(\left|Z_{t}\right|\right)\right)$, $\sigma_{t}^{2}$ is conditional variance, $\omega$, $\beta$, $\alpha$, $\theta$ and $\lambda$ are constant coefficients. $Z_{t}$ is a standard normal variable or comes from a student's-t distribution. Once volatility is calculated, then equation \ref{eq:var_covar} is used to predict VaR. 

\subsection{Semi-parametric Methods}

\subsubsection{CaviaR asymmetric}
Asymmetric  Conditional Autoregressive approach directly models VaR for return ${x_{t}} $ as follows
\begin{equation}\mathrm{VaR}_{t}=\beta_{1}+\beta_{2} \mathrm{VaR}_{t-1}+\beta_{3}\left(x_{t-1}\right)^{+}+\beta_{4}\left(x_{t-1}\right)^{-},\end{equation}
where $\beta_{i}$ are constants and $y^{+} = max(y,0)$ and $y^{-} = - min(y,0)$. The $\beta_{i}$ coefficients minimize the following function
\begin{equation}
\min _{\beta \in \mathbb{R}^{k}} \frac{1}{T}\Big\{\Big(\Theta-1(x_{t}<\operatorname{VaR}_{t})\Big)(x_{t}-\operatorname{VaR}_{t})\Big\}.
\end{equation}

\subsubsection{Extreme Value Theory}
As described earlier in this paper, this method deals with values which are above a certain threshold. In the unconditional EVT approach, one could select a proper threshold using various methods such as Hill plot, mean excess function and so on. After setting a rolling window and a suitable threshold, we can use equation \eqref{5} to calculate daily VaR.



\section{An Overview of Back-testing Methods}\label{sec:appendix_backtesting}

In this section, we present an overview of the back-testing methods used in our paper. As there are numerous back-testing methods proposed in the literature, we have employed three most popular of them to evaluate model performance from different perspectives. For model validation, we have implemented Kupiec and Christoffersen methods, and for comparing model predictability power with other competing models, we have used Diebold-Mariano predictive ability test.

\subsection{Kupiec test}
This test evaluates whether the number of realized violations are different from the predetermined violation rate. If $T$ is the number of observations and $x$ is the number of violations, under the null hypothesis we have
\begin{equation}
    H_{0}: p=\hat{p}=\frac{x}{T} ,
\end{equation}
where $\hat{p}$ is the realized violation rate, and $p$ is the violation rate corresponding to VaR quantile. This test is a likelihood-ratio test, where the test statistics is
\begin{equation}
    L R_{u c}=-2 \ln \left(\frac{(1-p)^{T-x} p^{x}}{\left[1-\left(\frac{x}{T}\right)\right]^{T-x}\left(\frac{x}{T}\right)^{x}}\right) ,
\end{equation}
under the null hypothesis, $L R_{uc}$  has a $\chi^{2}$ distribution with one degree of freedom. 

\subsection{Christoffersen test}
Christoffersen test is like Kupiec test, but in addition to the number of violations, it examines whether the violations are independent through time or not. For this purpose, an independent component is added to the Kupiec test statistics. The test statistics is

\begin{equation}
    L R_{c c}=-2 \ln \left(\frac{(1-\pi)^{n_{00}+n_{10}} \pi^{n_{01}+n_{11}}}{\left(1-\pi_{0}\right)^{n_{00}} \pi_{0}^ {n_{01}}\left(1-\pi_{1}\right)^{n_{10}} \pi_{1}^{n_{11}}}\right) +L R_{u c},
\end{equation}
where $n_{ij}$ is a variable that shows the number of periods when state $j$ occurred with respect to occurrence of state $i$ on the previous period. State 0 is a period where there is no violation, while state 1 is a period where there is a violation. Now $\pi_{i}$ is defined as the probability of observing a violation conditional on state $i$ on the previous period. Therefore, we have
\begin{equation}
\begin{aligned}
\pi_{0} =& \frac{n_{01}}{n_{00}+n_{01}},\\
\pi_{1} =& \frac{n_{11}}{n_{10}+n_{11}},\\
\pi =& \frac{n_{01}+n_{11}}{n_{00}+n_{01}+n_{10}+n_{11}}.
\end{aligned}
\end{equation}
Under the null hypothesis, $\pi_{0}$ and $\pi_{1}$ should be equal. $L R_{c c}$  has a $\chi^{2}$ distribution with two degrees of freedom. 

\subsection{Diebold-Mariano Predictive Ability test}
In this test, we compare only two methods at the same time. The null hypothesis of this framework assumes that the loss series generated by one of the forecasting methods is no worse than the other method. If we name the loss series of method $i$ by $e_{i}$ then $d = g(e_{i}) - g(e_{j})$ is the loss differential series of methods $i$ and $j$. $g$ is a loss function like $g(e_{i}) = e_{i}^{2}$. The test statistics is
\begin{equation}
    S_{2}=\sum_{i=1}^{T} I_{+}\left(d_{t}\right),
\end{equation}
where
\begin{equation}
 I_{+}\left(d_{t}\right)=
 \begin{cases}
1 \quad & \text { if } d_{t}>0, \\
0 \quad & \text { otherwise }.
\end{cases}
\end{equation}
Under the null hypothesis, test statistics has a binomial distribution with parameters $T$ and 0.5, where $T$ is the number of observations. As discussed by \cite{diebold2002comparing}, in large samples the test statistics becomes
\begin{equation}
    S_{2 a}=\frac{S_{2}-0.5 T}{\sqrt{0.25 T}} \stackrel{a}{\sim} \mathcal{N}(0,1).
\end{equation}

\end{document}